\documentclass[preprint,review,12pt]{elsarticle}

\usepackage{color}
\usepackage{graphicx}
\usepackage{upgreek}
\usepackage{epstopdf}
\usepackage{bm}
\usepackage{booktabs}
\usepackage{amsmath}
\usepackage{lineno}
\usepackage{hyperref}
\hypersetup{pdfauthor=author}
\usepackage{setspace}
\usepackage{geometry}
\usepackage[title,titletoc]{appendix} 
\usepackage{enumitem}
\usepackage{amssymb}
\usepackage{makecell}
\usepackage{caption}
\usepackage{tcolorbox}
\tcbuselibrary{listings}
\tcbuselibrary{breakable}
\tcbset{
    listing engine=listings,
    colback=white!95!blue,
    colframe=blue!75!black,
    listing options={
        basicstyle=\ttfamily,
        numbers=left,
        numberstyle=\tiny,
        stepnumber=1,
        showstringspaces=false,
        breaklines=true,
        breakatwhitespace=true,
        tabsize=4,
        language=C++
    }
}

\newcommand{\etal}{\textit{et al}.}

\let\oldequation\equation
\let\oldendequation\endequation

\renewenvironment{equation}
  {\linenomathNonumbers\oldequation}
  {\oldendequation\endlinenomath}

\biboptions{sort&compress}

\graphicspath{{Fig/}}

\begin{document}

\begin{frontmatter}

\title{
Fine-tuning a Large Language Model for Automating Computational Fluid Dynamics Simulations
}

\author[fir]{Zhehao Dong}

\author[fir]{Zhen Lu\corref{cor1}}
\ead{zhen.lu@pku.edu.cn}
\author[fir,sec]{Yue Yang\corref{cor1}}
\ead{yyg@pku.edu.cn}

\cortext[cor1]{Corresponding author.}

\address[fir]{State Key Laboratory for Turbulence and Complex Systems, College of Engineering, Peking University, Beijing 100871, China}
\address[sec]{HEDPS-CAPT, Peking University, Beijing 100871, China}

\begin{abstract}

Configuring computational fluid dynamics (CFD) simulations typically demands extensive domain expertise, limiting broader access.
Although large language models (LLMs) have advanced scientific computing, their use in automating CFD workflows is underdeveloped. 
We introduce a novel approach centered on domain-specific LLM adaptation. 
By fine-tuning Qwen2.5-7B-Instruct on NL2FOAM, our custom dataset of 28716 natural language-to-OpenFOAM configuration pairs with chain-of-thought (CoT) annotations enables direct translation from natural language descriptions to executable CFD setups. 
A multi-agent system orchestrates the process, autonomously verifying inputs, generating configurations, running simulations, and correcting errors. 
Evaluation on a benchmark of 21 diverse flow cases demonstrates state-of-the-art performance, achieving 88.7\% solution accuracy and 82.6\% first-attempt success rate. 
This significantly outperforms larger general-purpose models like Qwen2.5-72B-Instruct, DeepSeek-R1, and Llama3.3-70B-Instruct, while also requiring fewer correction iterations and maintaining high computational efficiency. 
The results highlight the critical role of domain-specific adaptation in deploying LLM assistants for complex engineering workflows.
Our code and fine-tuned model have been deposited at \url{https://github.com/YYgroup/AutoCFD}.

\end{abstract}



\end{frontmatter}

\newpage

\section{Introduction}

Computational fluid dynamics (CFD) has become an indispensable tool across aerospace~\cite{mani2023perspective}, energy~\cite{Ren2014Numerical}, and biomechanical~\cite{Lu2024GPU} applications, enabling the simulation of complex phenomena such as turbulence~\cite{Yang2023Applications}, multiphase flows~\cite{Hu2024Improving}, and combustion~\cite{Zhang2023Modeling}. 
Despite its widespread adoption, CFD remains inaccessible to many due to its steep learning curve, requiring expertise in numerical methods, programming, and domain-specific software like OpenFOAM~\cite{jasak2007openfoam}.
Manually preparing configuration files and debugging via command-line interfaces is error-prone and time-consuming. 
While graphical interfaces offer some relief, they still demand significant manual effort and specialized knowledge. 
Recent advances in large language models (LLMs) offer a transformative opportunity to automate these complex CFD workflows through natural language interaction, potentially democratizing access to this powerful tool.

LLMs demonstrate remarkable natural language capabilities across diverse domains~\cite{achiam2023gpt,qwen2024qwen2.5,guo2025deepseek,dubey2024llama}. 
They have aided scientific discovery~\cite{Birhane2023Science} in fields like mathematics~\cite{Azerbayev2024} and chemistry~\cite{Jablonka2024Leveraging}.
However, their application to computational physics, particularly CFD, is constrained by the need for precise physical understanding and software-specific syntax. 
While general-purpose LLMs like GPT-4~\cite{achiam2023gpt} and DeepSeek-R1~\cite{guo2025deepseek} contain broad scientific knowledge, they lack the specialized expertise crucial for reliable CFD automation, often generating physically inconsistent parameters or syntactically incorrect configurations. 
This domain knowledge gap hinders effective automation of the complex CFD workflow, which demands a deep understanding of numerical schemes, turbulence models, boundary conditions, and solver-specific implementation details. 
Addressing this challenge requires domain-specific adaptation that aligns LLM capabilities with the rigorous requirements of CFD.

Retrieval-augmented generation (RAG)~\cite{lewis2020retrieval} has been proposed to address the domain knowledge gap in CFD automation by allowing general-purpose LLMs to access specialized information.
Examples include MetaOpenFOAM~\cite{chen2024metaopenfoam}, a RAG-based multi-agent system that automates OpenFOAM simulation workflows from natural language inputs and later extended to post-processing~\cite{Chen2025MetaOpenFOAM}.
Similarly, Pandey \etal~\cite{Pandey2025OpenFOAMGPTa} developed RAG-based OpenFOAMGPT, demonstrating zero-shot case setup and condition modification capabilities with GPT-4o and GPT-o1.
The OpenFOAMGPT framework was then utilized to evaluate the cost-effectiveness of different LLMs~\cite{Wang2025status}.
However, RAG-based systems are inherently limited by their reliance on knowledge retrieval; they access pre-existing information rather than developing true domain understanding~\cite{siriwardhana2023improving,zhang2024raft}. 
This limitation becomes particularly problematic for new configurations or complex physical scenarios absent from reference materials, potentially leading to fragmented knowledge integration and physically inconsistent outputs~\cite{gao2024retrievalaugmented,lyu2024curd}. 
Furthermore, selecting optimal numerical configurations in CFD is challenging as settings are often non-unique and highly dependent on specific flow regimes and geometric complexities---a nuance difficult to capture through simple retrieval.
These shortcomings underscore the need to embed domain expertise directly into model parameters.

Fine-tuning~\cite{Ziegler2020Finetuning} offers a direct approach for embedding domain expertise into LLMs, potentially overcoming RAG's limitations by incorporating CFD knowledge into the model's parameters rather than relying on external retrieval. 
It enables models to develop a deeper understanding of fluid dynamics principles, numerical methods, and solver-specific requirements. 
Effective fine-tuning requires high-quality labeled datasets~\cite{Ouyang2022Training} specifically tailored to CFD applications---a challenge given the complexity and diversity of simulation scenarios. 
Such datasets must capture not only correct syntax and parameter settings but also the underlying physical reasoning and problem-specific considerations that guide expert decisions. 
Despite the challenges, fine-tuned models offer potentially greater consistency in handling new cases, enhanced physics reasoning, and robust performance without the computational overhead and latency associated with retrieval systems~\cite{Wei2021Finetuned}. 

In this work, we developed a domain-specific fine-tuned LLM for automating CFD workflows. 
We developed NL2FOAM, a custom dataset comprising 28716 pairs of natural language descriptions and corresponding OpenFOAM configurations, augmented with chain-of-thought (CoT) annotations to capture expert reasoning. 
Using NL2FOAM, we fine-tuned Qwen2.5-7B-Instruct~\cite{qwen2024qwen2.5}, enabling it to translate high-level natural language problem descriptions into executable CFD setups. 
A multi-agent system manages the workflow, handling input verification, configuration generation, simulation execution, and error correction autonomously. 
Our evaluation on a benchmark of 21 diverse flow cases demonstrates state-of-the-art performance, significantly surpassing larger general-purpose models and highlighting the effectiveness of specialized fine-tuning for complex engineering tasks.

The remainder of this paper is organized as follows. 
Section~\ref{sec:method} introduces our methodological framework, detailing the fine-tuning approach, multi-agent system architecture, NL2FOAM dataset construction, and benchmark setup. 
Section~\ref{sec:result} presents comprehensive validation results, comparing our fine-tuned model against general-purpose LLMs across multiple metrics, including an ablation study on CoT reasoning. 
Finally, Section~\ref{sec:conclusion} summarizes our findings, discusses limitations, and outlines directions for future research in LLM-assisted CFD automation.

\section{Methodology}
\label{sec:method}

\subsection{Fine-tuning LLM}

LLMs, pre-trained on vast corpora of text, encode broad knowledge and language capabilities~\cite{achiam2023gpt,guo2025deepseek,qwen2024qwen2.5,dubey2024llama}.
They can be specialized for domain-specific applications through fine-tuning~\cite{Ouyang2022Training}, particularly when sufficient labeled training data exists and high-precision is essential. 
Conventional fine-tuning updates all model parameters, imposing significant computational burden. 
Low-rank adaptation (LoRA)~\cite{hu2022lora} substantially reduces the computational footprint while maintaining comparable performance. 
Mathematically, LoRA updates a pre-trained weight matrix $W \in \mathbb{R}^{d \times k}$ by adding $\delta W = BA$ with $B \in \mathbb{R}^{d \times r}$, $A \in \mathbb{R}^{r \times k}$, and the rank $r \ll \min(d,k)$. 
This approach typically reduces tunable parameters by 100- to 1000-fold.
During inference, the updated weight $W' = W + \delta W$ is used efficiently without additional computational overhead. 

For the CFD-specific adaptation, we fine-tuned Qwen2.5-7B-Instruct~\cite{qwen2024qwen2.5} via LoRA on the NL2FOAM dataset, which comprises 28716 natural language-to-OpenFOAM configuration cases. 
Details of NL2FOAM will be introduced in Sec.~\ref{sec:dataset},
and fine-tuning specifics are provided in~\ref{sec:appendix}.
In practice, the model processes a natural language description (e.g., ``Conduct a RANS simulation using k-epsilon model for a cylinder wake with inlet velocity of 10 m/s''), mesh files, and input file templates.
The mesh files are required as LLMs cannot reliably generate them, while templates are generated automatically from the description and mesh files. 
Although the fine-tuned model generally infers appropriate initial flow fields, explicit specification in the description improves success rates. 
The resulting fine-tuned LLM functions as an intelligent interface, translating user descriptions into executable OpenFOAM configurations without requiring expertise in its syntax and parameter structures.

\subsection{Multi-agent system}
\label{sec:multi-agent-system}

\begin{figure}[ht]
\centering
\includegraphics{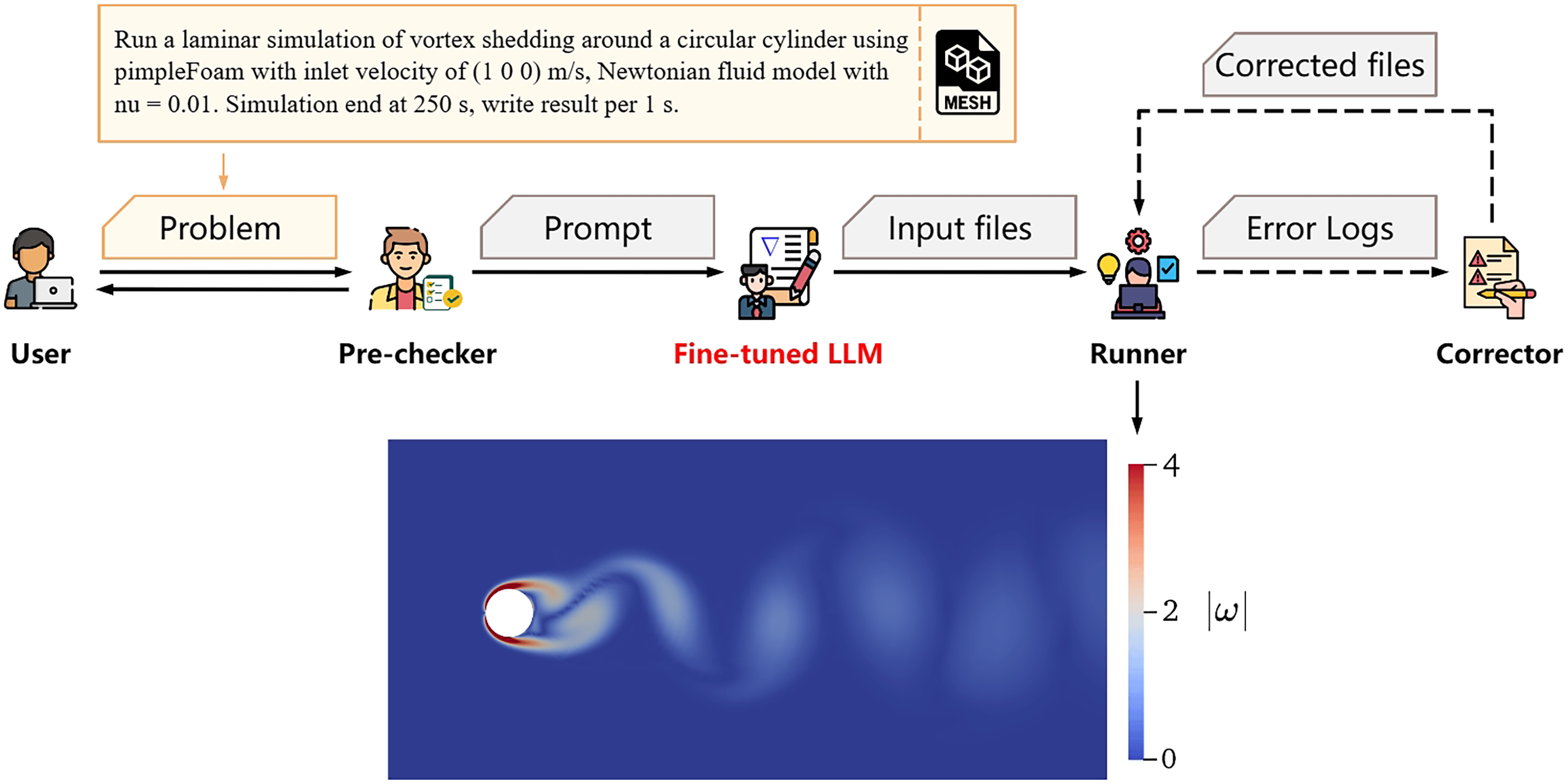}
\caption{
The multi-agent system for automated CFD simulation, centered on a fine-tuned LLM. The workflow progresses from user input through a pre-checker, LLM-based generation of input files, simulation execution by the runner, and an iterative error correction loop involving the corrector, ultimately producing simulation results.
}
\label{fig:multi-agent}
\end{figure}

The fine-tuned LLM serves as the core reasoning engine in our multi-agent system. 
This system orchestrates CFD workflow automation and enhances domain-specific reasoning, building upon previous RAG approaches~\cite{chen2024metaopenfoam}.
As illustrated in Fig.~\ref{fig:multi-agent}, this system orchestrates four specialized agents---\emph{pre-checker}, \emph{LLM generator}, \emph{runner}, and \emph{corrector}---through a structured workflow that enforces OpenFOAM syntax compliance and numerical stability. Further details on the multi-agent system setup can be found in the supplementary material.

The multi-agent workflow begins with a user's natural language description of the CFD problem and mesh files.
The pre-checker validates inputs, queries users if needed, and generates input templates incorporating boundary names extracted from the mesh files.
The fine-tuned LLM then generates the OpenFOAM case directory through structured CoT reasoning, including numerical configurations, initial fields, boundary conditions, and an execution script.
The runner executes simulations while monitoring real-time logs.
If errors occur, the corrector analyzes and resolves issues.
The corrected files are then resubmitted to the runner, continuing this cycle until the simulation completes successfully.

\subsection{NL2FOAM}
\label{sec:dataset}

\begin{figure}[ht]
\centering
\includegraphics{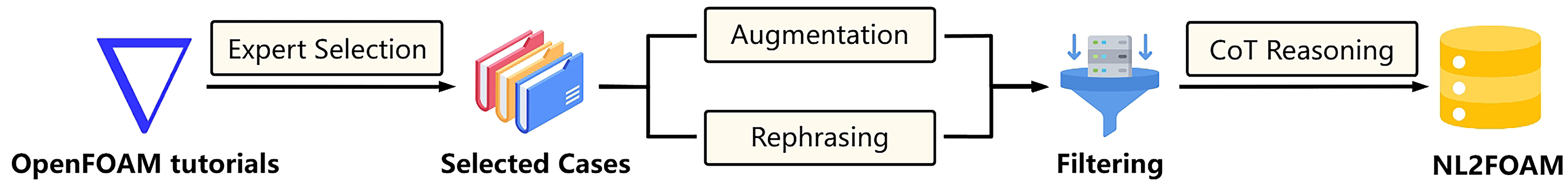}
\caption{
NL2FOAM construction pipeline. From 16 curated OpenFOAM cases, automated tools modify input files (\texttt{controlDict}, \texttt{fvScheme}, and \texttt{fvSolution}) to create 100k variations, while an LLM enhances linguistic diversity through description rephrasing.
Simulation testing eliminates cases with errors, solution divergence, or excessive runtime, followed by CoT reasoning to structure solutions.
}
\label{fig:pipeline}
\end{figure}

As sketched in Fig.~\ref{fig:pipeline}, we built the NL2FOAM dataset to fine-tune an LLM for CFD automation. 
Starting with 16 OpenFOAM cases (see Tab.~\ref{tab:nl2foam}) spanning laminar and turbulent flows with the Reynolds number ($Re$) from $10$ to $9\times 10^7$, we modified configuration files to generate over 100k variations. 
An LLM enhanced linguistic diversity by rephrasing the problem descriptions.
Test simulations filtered out cases with runtime errors, solution divergence, or excessive runtime (more than 72 hours).
The final dataset contains 28716 cases that link natural language descriptions to executable OpenFOAM configurations, each including a problem description, mesh files, OpenFOAM input files (numerical configurations, initial fields, and boundary conditions), an execution script, and an LLM-generated CoT reasoning trace.

\begin{table}[!ht]
    \centering
    \caption{Selected OpenFOAM cases used to build the NL2FOAM dataset.}
    \label{tab:nl2foam}
    \begin{tabular}{lll}
    \toprule
        \textbf{Case name} & $ \bm{Re} $ & \textbf{Solver} \\ 
    \midrule
        Cavity & $10 \sim 1 \times 10^{4}$ & \makecell[l]{icoFoam\\ pisoFoam} \\ \hline
        Cylinder wake & 100 & icoFoam \\ \hline
        Elbow & 1200 & icoFoam \\ \hline
        Poiseuille flow & 1500 & icoFoam \\ \hline
        Planar poiseuille flow & $360 \sim 460$ & pimpleFoam \\ \hline
        Backward-facing step & $2.5 \times 10^{4}$ & \makecell[l]{pisoFoam \\ simpleFoam \\ pimpleFoam} \\ \hline
        TJunction & 2600 & pimpleFoam \\ \hline
        Airfoil2D & $9 \times 10^{7}$ & simpleFoam \\ \hline
        Mixervessel2D & $5 \times 10^{6}$ & simpleFoam \\ \hline
        MixervesselAMI2D & $1.5 \times 10^{5}$ & pimpleFoam \\ \hline
        Rotating cylinders & 200 & simpleFoam \\ 
    \bottomrule
    \end{tabular}
\end{table}

The NL2FOAM dataset exhibits diversity across multiple dimensions. 
We selected various OpenFOAM cases covering steady and unsteady flows, laminar and turbulent regimes, and multiple simulation methods (directly solving the Navier-Stokes equations for laminar flows, and Reynolds-averaged Navier-Stokes simulations and large-eddy simulations for turbulent flows).
We further expanded diversity through data augmentation, varying numerical parameters including finite volume schemes, velocity-pressure coupling solvers, solver parameters, and linear system solvers. 
Additionally, LLM-based rephrasing of problem descriptions ensures linguistic diversity in the natural language specifications. 
Quality is assured through test simulations, eliminating invalid configurations from the final dataset.

Despite these strengths, the dataset is limited to incompressible flow cases, excluding more complex phenomena such as multiphase, compressible, or reacting flows, which often require additional fields and distinct solver settings.
Additionally, the limited number of base OpenFOAM cases restricts coverage even within incompressible flows. 
In the present work, we limited our initial scope to incompressible flows to establish a robust foundation.
Future work should systematically expand to include more complex flows, incorporating domain-specific constraints.

Each training case in NL2FOAM is a string pairing a structured input with a ground truth response, as illustrated in Fig.~\ref{fig:NL2FOAM}.
The input includes a natural language problem description, mesh file references, and input templates. 
The ground truth contains complete OpenFOAM configurations and an execution script, preceded by CoT reasoning steps.  
We utilized the synthetic CoT solutions~\cite{wei2022chain} to enhance small model performance~\cite{yu2024metamath}.
These ground truth CoT annotations are generated by a separate LLM, following a six-step reasoning structure provided in the prompt: (1) define the problem and objectives; (2) select the appropriate solver and turbulence models; (3) determine the required OpenFOAM files; (4) apply the boundary and initial conditions; (5) configure the simulation parameters; (6) write the execution script. 
The final OpenFOAM files and script adhere to this reasoning.
\begin{figure}[htbp]
\centering
    \begin{tcolorbox}[colback=blue!5!white,colframe=blue!5!black, title=An example case in NL2FOAM]
\# Question:\\
Conduct a RANS simulation for the turbulent flow over a backward-facing step using the pimpleFoam solver. Set the inlet velocity to 10 m/s and the outlet pressure to 0. Use the k-epsilon turbulence model for this simulation. The fluid is Newtonian with a kinematic viscosity of $1 \times 10\textsuperscript{-5}$ m\textsuperscript{2}/s. Simulation end at 0.04 s, write result per 0.005 s.

\# Mesh file: \texttt{blockMeshDict}\\
\# Input file template:
\begin{verbatim}
{
    "0/p": {
        ...
        "boundaryField": {
            "inlet": { "type": "xxx", ... },
            ...
        }
    },
    ...
}
\end{verbatim}

\tcblower

\# Response:\\
\# CoT reasoning steps:\\
\#\# 1. define the problem and objectives: ...\\
\#\# 2. select the appropriate solver and turbulence models: ...\\
\#\# 3. determine the required OpenFOAM files: ...\\
\#\# 4. apply the boundary and initial conditions: ...\\
\#\# 5. configure the simulation parameters: ...\\
\#\# 6. write the execution script: ...\\
\# Based on the above thought process, the complete foam files and execution scripts are as follows:\\
\#\# \texttt{OpenFOAM files}\\
\#\# \texttt{execution script}
\end{tcolorbox}
\caption{
Structure of an example case within the NL2FOAM dataset used for fine-tuning. 
The input consists of a natural language description (Question), mesh file references, and input templates.
The output (Response) includes CoT reasoning steps followed by the generated OpenFOAM files and execution script.
}
\label{fig:NL2FOAM}
\end{figure}

\subsection{Benchmark}

We evaluate our framework using an incompressible flow benchmark of 10 laminar and 11 turbulent cases with $Re$ spanning from $40$ to $5\times 10^8$, as listed in Tab.~\ref{tab:nl2foam-bench}. 
71\% of test cases (15/21) extend beyond the OpenFOAM tutorial, including a jet flow and turbulent square column wakes.
To ensure a fair assessment, there is no configuration overlap between the training and benchmark sets. 
Although two case names appear in both sets (cylinder wake and Poiseuille flow), their parameters differ substantially.
While the training set includes a laminar cylinder wake at $Re=100$, the benchmark tests this geometry from $Re=40$ to $1\times 10^5$ (excluding 100). 
Likewise, the Poiseuille flow parameters change from $Re=1500$ (training) to $Re=100$ (benchmark). 
This separation ensures the evaluation assesses generalization across diverse $Re$ and flow regimes, not memorization. 
Furthermore, the benchmark includes multi-solver configurations (e.g., cylinder wake validated with icoFoam, simpleFoam, pisoFoam, and pimpleFoam) to test the framework's ability to select context-appropriate numerical methods.
\begin{table}[!ht]
    \centering
    \caption{
    Benchmark cases used for evaluating the LLM-based CFD automation framework, comprising 21 diverse flow scenarios across $Re$ from 40 to $5\times 10^8$. 
    Cases marked with $\bigcirc$ in the tutorial column are from the OpenFOAM tutorials, while the remaining 71\% are variations or distinct problems designed to assess the generalization capabilities.
    }
    \label{tab:nl2foam-bench}
    \begin{tabular}{lllc}
    \toprule
        \textbf{Case name} & $\bm{Re}$ & \textbf{Solver} & \textbf{Tutorial} \\ 
    \midrule
        Poiseuille flow & 100 & \makecell[l]{icoFoam\\ simpleFoam} & \\ \hline
        Square column wake & $5 \times 10^{5} \sim 5 \times 10^{8}$ & pimpleFoam &  \\ \hline
        Cylinder wake & $40 \sim 1 \times 10^{5}$ & \makecell[l]{icoFoam\\ simpleFoam\\ pisoFoam\\ pimpleFoam} &  \\ \hline
        Jet & $3.3 \times 10^{5}$ & simpleFoam &  \\ \hline
        Couette flow & $66 \sim 6.6 \times 10^{4}$ & pimpleFoam & $\bigcirc$ \\ \hline
        Square bend & $200 \sim 2 \times 10^{4}$ & simpleFoam &  $\bigcirc$ \\ \hline
        Forward-facing step & $50 \sim 5 \times 10^{4}$ & simpleFoam & $\bigcirc$ \\ 
    \bottomrule
    \end{tabular}
\end{table}

We evaluated performance using five metrics: accuracy, pass@1, iterations, token usage, and expense. 
An ``experiment'' proceeds from inputting a natural language description and mesh files to obtaining CFD simulation results. 
An experiment ``passes'' if it achieves a convergent solution in 72 hours at most 10 correction attempts; otherwise, it ``fails''. 
Each benchmark case undergoes $n=10$ independent experiments. 
The final reported metrics are averaged across all experiments.

``Accuracy'' measures solution reliability using the L2 norm $\epsilon$ between the LLM-based automated CFD solution and the benchmark, defined as $1-\epsilon$. Failed experiments receive 0\% accuracy.
``pass@$1$''~\cite{chen2021evaluating} represents the first-attempt success rate, defined as the ratio of the number of successful experiments to the total number of attempts (10 attempts in this study). 
``Iterations'' count the correction rounds needed to fulfill user requirements, indicating convergence efficiency. 
``Token Usage'' measures LLM tokens consumed, reflecting computational resource requirements. 
``Expense'' quantifies actual costs per experiment, differing from token usage due to varying prices.

\section{Results}
\label{sec:result}

We evaluated our fine-tuned LLM against open-source LLMs (Qwen2.5-72B-Instruct~\cite{qwen2024qwen2.5}, DeepSeek-R1~\cite{guo2025deepseek}, and Llama3.3-70B-Instruct~\cite{dubey2024llama}) and the RAG-based MetaOpenFOAM~\cite{chen2024metaopenfoam} (using GPT-4o~\cite{achiam2023gpt}). 
To isolate the impact of base LLM performance, we integrated the open-source LLMs into our framework through component substitution, retaining the multi-agent architecture’s verification and error-correction modules. 
The sampling randomness parameter, known as ``temperature'' for LLMs, was set to 0.7 for these open-source models. 
For the MetaOpenFOAM comparison, we used the same mesh files and adopted its GPT-4o temperature setting of 0.01. 

\subsection{Overall Performance}

\begin{figure}[ht]
  \centering
  \includegraphics{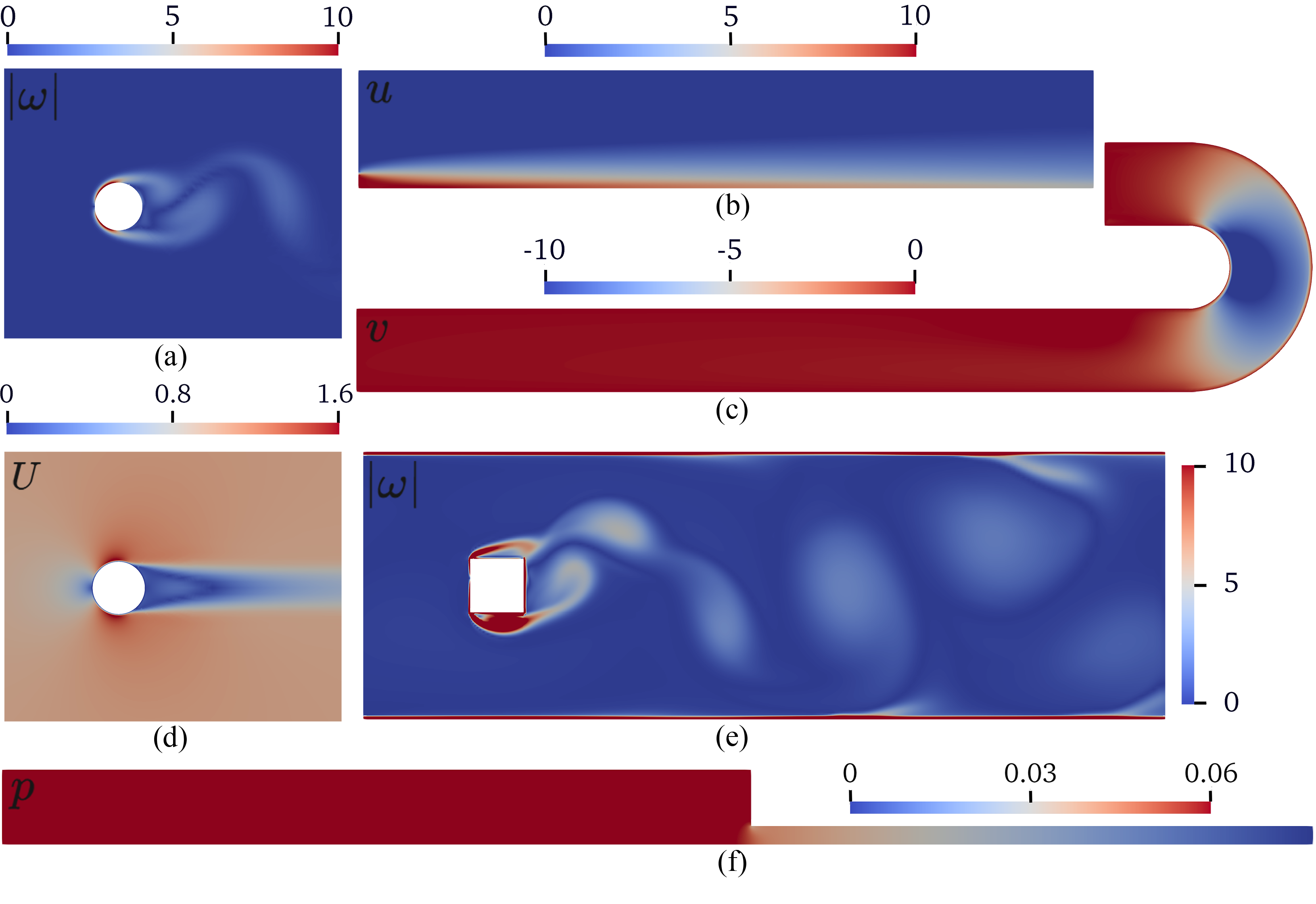}
  \caption{
  Simulation results from benchmark cases generated using our automation workflow based on the fine-tuned LLM: (a) vorticity magnitude $\vert\bm{\omega}\vert$ for cylinder wake (\(Re = 200\)), (b) velocity component $u$ for jet flow (\(Re = 3.3 \times 10^5\)), (c) velocity component $v$ for square bend (\(Re = 2 \times 10^4\)), (d) velocity magnitude $U$ for cylinder wake (\(Re = 1 \times 10^5\)), (e) voricity magnitude $\vert\bm{\omega}\vert$ for square column wake in a channel (\(Re = 5 \times 10^5\)), and (f) pressure field $p$ for forward-facing step (\(Re = 50\)).
  }
  \label{fig:visualization}
\end{figure}

Our fine-tuned LLM demonstrates robust performance across benchmarks, achieving 88.7\% accuracy and 82.6\% pass@1 with simulation reliability and physical consistency. 
Figure~\ref{fig:visualization} confirms the method generates correct OpenFOAM configurations for diverse cases. 
The visualizations show that the obtained velocity distributions and vortex shedding patterns align with established CFD benchmarks. 
The results capture essential flow phenomena including vortex shedding behind obstacles in Figs.~\ref{fig:visualization}a and e, jet diffusion in Fig.~\ref{fig:visualization}b, characteristic bend flows in Fig.~\ref{fig:visualization}c, and averaged wake obtained in RANS in Fig.~\ref{fig:visualization}d. 
This validation confirms our method effectively automates CFD configuration generation across laminar and turbulent regimes while maintaining adherence to physical principles, bridging LLM-driven automation and simulation requirements.

\begin{figure}[ht]
\centering
\includegraphics{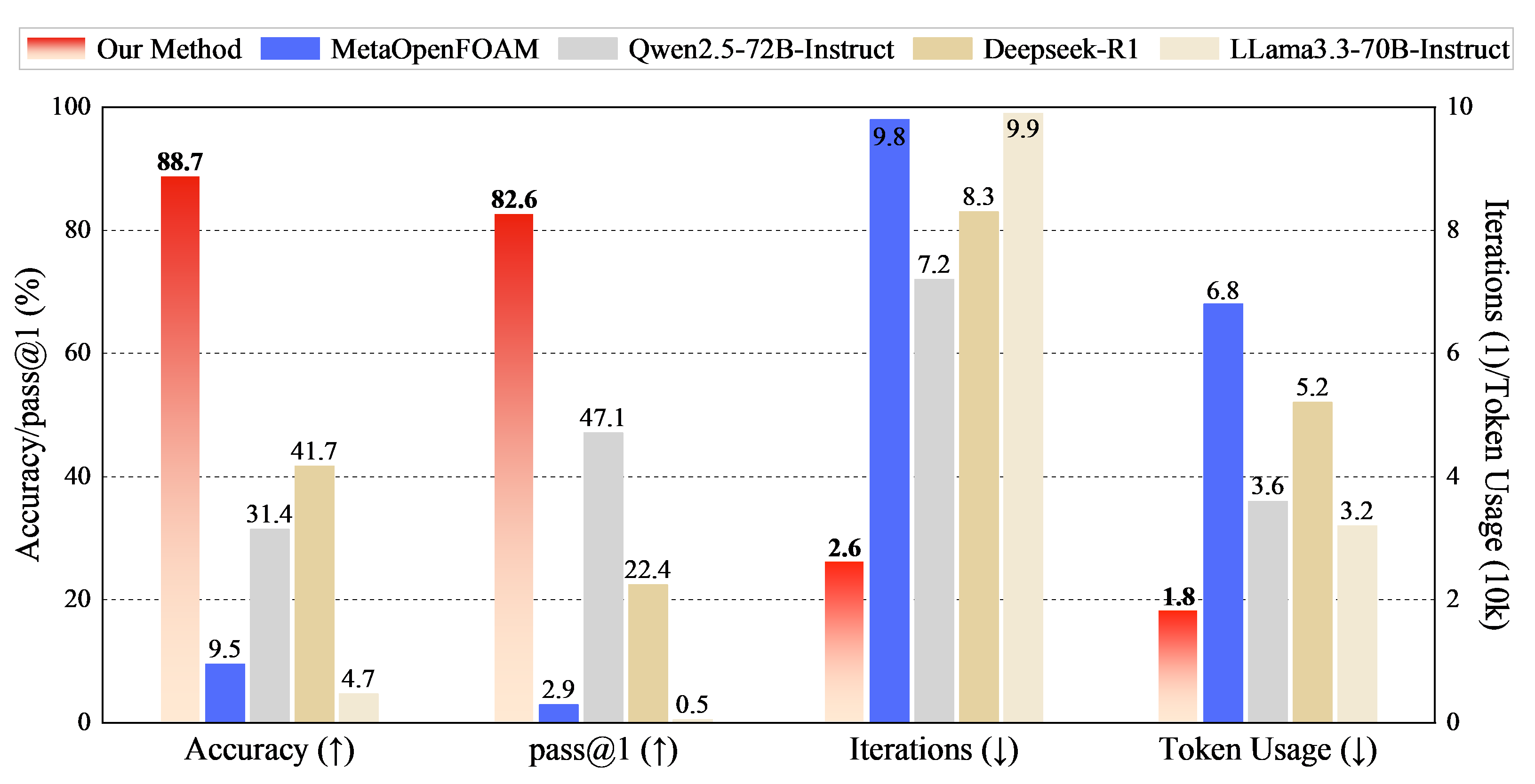}
\caption{Benchmark performance of different methods, comparing our method against four baselines (MetaOpenFOAM, Qwen2.5-72B-Instruct, Deepseek-R1, and LLama3.3-70B-Instruct) across accuracy, pass@1, iterations, and token usage. 
}
\label{fig:baseline}
\end{figure}

Comparative benchmarking in Fig.~\ref{fig:baseline} establishes our approach as state-of-the-art in both solution quality and operational efficiency. 
Our method leads significantly across all metrics: 88.7\% accuracy (vs.~41.7\% by Deepseek-R1), 82.6\% pass@1 (surpassing runner-up Qwen2.5-72B-Instruct by 35.5\%), and requiring only 2.6 correction iterations (vs.~7.2 by Qwen2.5-72B-Instruct) with 1.8k token usage (vs.~3.2k by Llama3.3-70B-Instruct).
This enhanced performance stems from fine-tuning the LLM on NL2FOAM, which improves its grasp of physics principles and OpenFOAM syntax, thereby reducing errors. 
In contrast, lower-performing methods like Llama3.3-70B-Instruct and MetaOpenFOAM only succeed with the Poiseuille flow problem, with MetaOpenFOAM possibly limited by its RAG knowledge library.
Furthermore, our model is highly cost-effective. Its token efficiency (17816 tokens/case) leads to a low cost at 0.020 USD per solution, compared to 0.035 USD for Qwen2.5-72B-Instruct, 0.042 USD for DeepSeek-R1, 0.018 USD for Llama3.3-70B-Instruct, and 0.227 USD for MetaOpenFOAM. 

Benchmark demonstrates that larger general-purpose models, lacking specialized CFD knowledge, frequently generate incomplete or inadequate configurations. 
Our analysis showed they often omit necessary files or parameters (e.g., solver-specific settings) and produce parameter values lacking the diversity and flexibility needed for varied CFD scenarios. 
This inflexibility, particularly with different turbulence models or flow conditions, results in higher simulation failure or divergence rates.

In contrast, domain-specific fine-tuning enables the model to effectively bridge natural language instructions and the OpenFOAM setup.
This results in more relevant and robust configurations tailored to CFD tasks, confirmed by the lower iteration metric.
Tracking correction iterations revealed that most errors involved missing basic parameters, such as a pressure reference, which the corrector agent resolved efficiently. 
The most persistent challenges occur in turbulence simulations, mainly due to divergence caused by unreasonable parameter values, such as dissipation rates in the $k$-$\omega$ model was set orders of magnitude too high.
Future work will extend the dataset with more diverse examples to improve fine-tuning and reduce these failures.

\subsection{Ablation study}
\label{sec:ablation}

\begin{figure}[ht]
\centering
\includegraphics{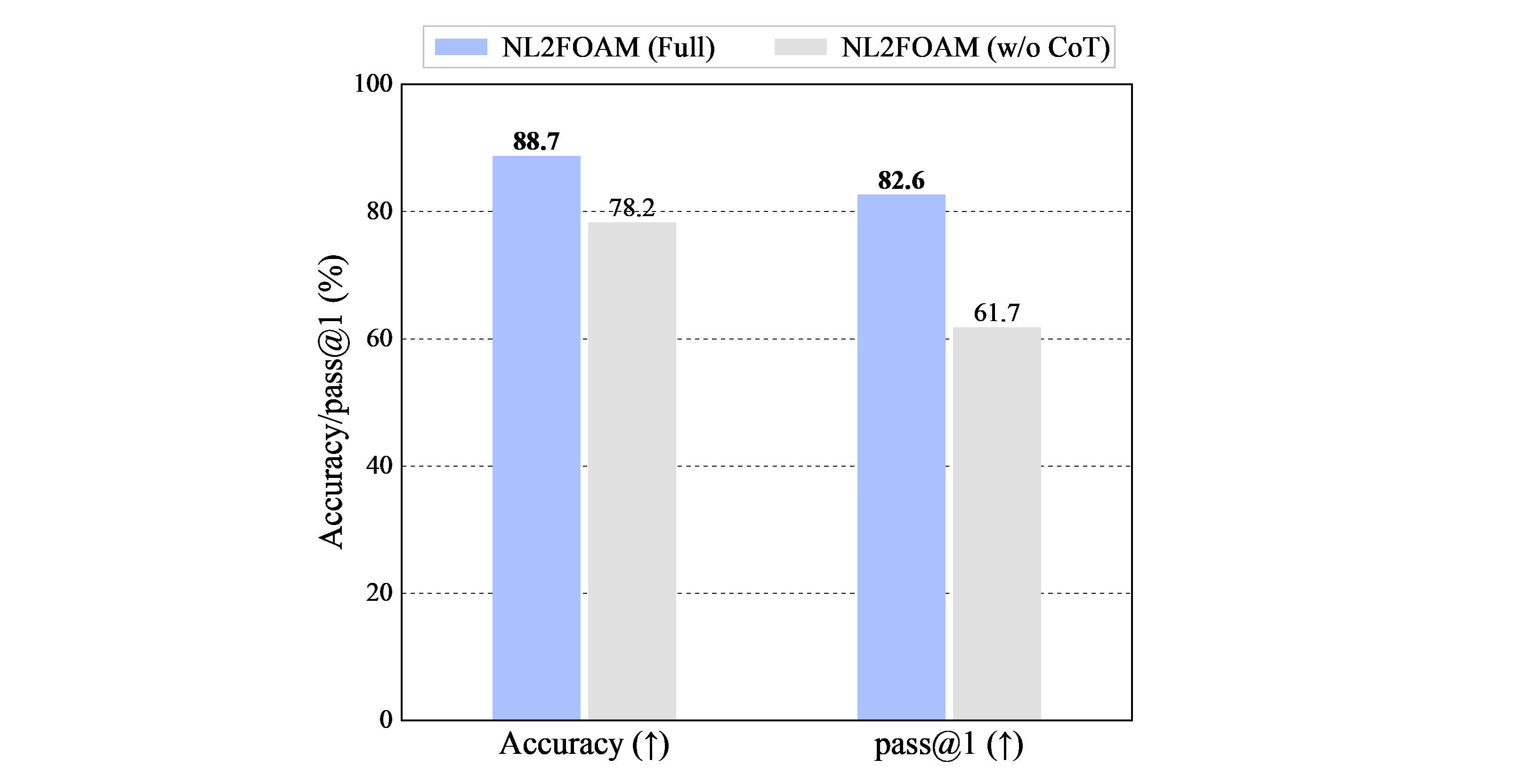}
\caption{
Ablation study on the impact of CoT reasoning. 
The chart compares the accuracy and first-attempt success rate (pass@1) of the fine-tuned LLM trained with the full NL2FOAM dataset (including CoT, blue bars) versus a dataset variant without CoT annotations (grey bars).
}
\label{fig:ablation-cot}
\end{figure}

We quantified the impact of explicit CoT reasoning through an ablation study. 
Reconstructing the NL2FOAM dataset without CoT reasoning while preserving identical OpenFOAM inputs and execution scripts resulted in significant performance degradation. 
As Fig.~\ref{fig:ablation-cot} shows, the full dataset achieved 88.7\% accuracy and 82.6\% pass@1, improvements of 10.5\% and 20.9\% respectively over the CoT-ablated baseline (78.2\% accuracy and 61.7\% pass@1). 
This aligns with expectations that intermediate reasoning steps are crucial for complex physical modeling, where parameter selection requires deliberate computation rather than just pattern-based generation.
Error analysis indicated that models without CoT reasoning struggled with generating appropriate initial conditions and produced redundant settings. 
Conversely, the CoT-enhanced model reduced these errors, showing improved internalization of parameter configurations and physical constraints, with the performance gap widening for tasks requiring multi-step reasoning.

\section{Conclusion}
\label{sec:conclusion}

We developed an approach to automate CFD simulations by fine-tuning an LLM on domain-specific data. 
We constructed NL2FOAM, a dataset of 28716 natural language-to-OpenFOAM configuration pairs with chain-of-thought reasoning annotations, and fine-tuned Qwen2.5-7B-Instruct using LoRA to reduce trainable parameters from 7.6B to 0.02B. 
This domain-specific adaptation enables the LLM to translate natural language descriptions into complete OpenFOAM configurations. 
This fine-tuned LLM serves as the core of a multi-agent system (including pre-checker, LLM generator, runner, and corrector agents) that orchestrates the simulation workflow, ensuring syntax compliance and numerical stability.

Our approach achieved state-of-the-art performance with 88.7\% accuracy and 82.6\% pass@1 on a benchmark of 21 diverse cases spanning $Re$ from 40 to $5\times 10^8$. 
It significantly outperformed larger general-purpose models including Qwen2.5-72B-Instruct (31.4\% accuracy and 47.1\% pass@1), DeepSeek-R1 (41.7\% accuracy and 22.4\% pass@1), and Llama3.3-70B-Instruct (4.7\% accuracy and 0.5\% pass@1). 
Furthermore, our method required fewer correction iterations (2.6 vs.~7.2 for the runner-up) and achieved high token efficiency (17816 tokens/case), resulting in a low average cost of 0.020 USD per simulation. 
An ablation study confirmed that including CoT reasoning boosted accuracy by 10.5\% and pass@1 by 20.9\%, highlighting its value for complex physics simulations.

This research introduces a new paradigm for engineering automation that bridges natural language interfaces with specialized numerical simulations. 
By allowing non-experts to configure CFD simulations through natural language descriptions, our approach democratizes access to simulation capabilities while maintaining high accuracy. 
The multi-agent system further demonstrates how domain-specific LLMs can be integrated with verification and correction mechanisms to achieve reliable automation of technically complex workflows requiring both linguistic understanding and numerical precision.

While our method performs well on incompressible benchmarks, limitations remain for more complex simulations.
Future work will focus on expanding the NL2FOAM dataset to include more complex transport phenomena, e.g., multiphase flows, compressible flows, and heat transfer problems,
which involve additional configuration files and different solver setups.
Computational efficiency and accuracy should be assessed when scaling to industrial-level problems for practical application.
Furthermore, we plan to explore fine-tuning larger base models to enhance robustness and handle a wider variety of cases. 
These advancements will further broaden the approach's applicability for engineering challenges.

\section*{Acknowledgments}

This work has been supported in part by the National Natural Science Foundation of China (Nos.~52306126, 22350710788, 12432010, 11988102, and 92270203) and the Xplore Prize.

\appendix
\setcounter{figure}{0}
\setcounter{table}{0}

\section{Fine-tuning details}
\label{sec:appendix}

We applied LoRA with a rank $r=8$, reducing trainable parameters from 7.6B to 0.02B. 
Fine-tuning utilized Llamma-Factory~\cite{zheng-etal-2024-llamafactory}, a unified framework enabling efficient and flexible LLM adaptation. 
We trained the model using four NVIDIA GeForce RTX 4090 GPUs, employing AdamW~\cite{Loshchilov2019Decoupled} optimization.
The baseline learning rate was \(5 \times 10^{-5}\), with linear warmup over the first 10\% of steps to mitigate initial instability. 
A total batch size of 16 balanced GPU memory constraints and training efficiency.

\begin{figure}[ht]
\centering
\includegraphics{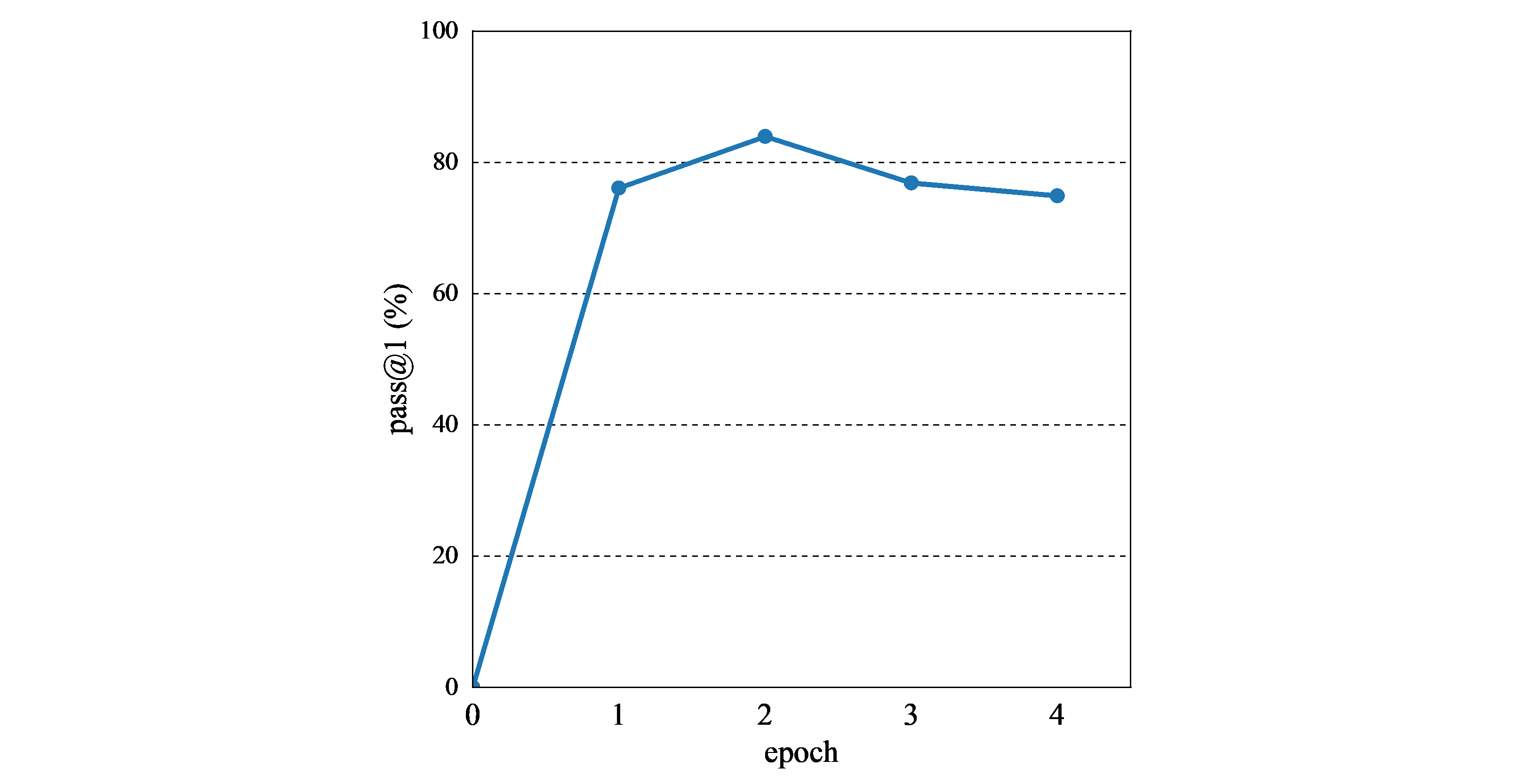}
\caption{
Evolution of the performance pass@1 across epochs on our benchmark. 
The model reached a peak pass@1 of 82.6\% at epoch two.
}
\label{fig:epochs}
\end{figure}

Training proceeded for four epochs, with checkpoints saved after each. 
This duration was empirically chosen based on validation loss plateaus to balance convergence and cost. 
As shown in Fig.~\ref{fig:epochs}, performance pass@1 peaked at 82.6\% after two epoches, then declined slightly while remained above 75\%, suggesting potential overfitting. 
Consequently, we selected the epoch two checkpoint for experiments.

\bibliographystyle{elsarticle-num}
\bibliography{autoCFD}

\begin{thebibliography}{10}
\expandafter\ifx\csname url\endcsname\relax
  \def\url#1{\texttt{#1}}\fi
\expandafter\ifx\csname urlprefix\endcsname\relax\def\urlprefix{URL }\fi
\expandafter\ifx\csname href\endcsname\relax
  \def\href#1#2{#2} \def\path#1{#1}\fi

\bibitem{mani2023perspective}
M.~Mani, A.~J. Dorgan, A perspective on the state of aerospace computational
  fluid dynamics technology, Annu. Rev. Fluid. Mech. 55 (2023) 431--457.

\bibitem{Ren2014Numerical}
Z.~Ren, Z.~Lu, L.~Hou, L.~Lu, Numerical simulation of turbulent combustion:
  Scientific challenges, Sci. China Phys, Mech. Astron. 57 (2014) 1495--1503.

\bibitem{Lu2024GPU}
Y.~Lu, P.~Wu, M.~Liu, C.~Zhu, A {GPU}-accelerated {3D ISPH-TLSPH} framework for
  patient-specific simulations of cardiovascular fluid–structure
  interactions, Comput. Methods Appl. Mech. Eng. 428 (2024) 117110.

\bibitem{Yang2023Applications}
Y.~Yang, S.~Xiong, Z.~Lu, Applications of the vortex-surface field to flow
  visualization, modelling and simulation, Flow 3 (2023) E33.

\bibitem{Hu2024Improving}
J.~Hu, Z.~Lu, Y.~Yang, Improving prediction of preferential concentration in
  particle-laden turbulence using the neural-network interpolation, Phys. Rev.
  Fluids 9 (2024) 34606.

\bibitem{Zhang2023Modeling}
S.~Zhang, Z.~Lu, Y.~Yang, Modeling the boundary-layer flashback of premixed
  hydrogen-enriched swirling flames at high pressures, Combust. Flame 255
  (2023) 112900.

\bibitem{jasak2007openfoam}
H.~Jasak, A.~Jemcov, Z.~Tukovic, et~al., {OpenFOAM}: A {C++} library for
  complex physics simulations, in: International workshop on coupled methods in
  numerical dynamics, Dubrovnik, Croatia, September 19-21, 2007.

\bibitem{achiam2023gpt}
OpenAI, {GPT-4} technical report (2023).
\newblock \href {http://arxiv.org/abs/2303.08774} {\path{arXiv:2303.08774}}.

\bibitem{qwen2024qwen2.5}
{Qwen Team}, Qwen2.5: A party of foundation models (2024).
\newblock \href {http://arxiv.org/abs/2412.15115} {\path{arXiv:2412.15115}}.

\bibitem{guo2025deepseek}
DeepSeek-AI, {DeepSeek-R1}: Incentivizing reasoning capability in {LLMs} via
  reinforcement learning (2025).
\newblock \href {http://arxiv.org/abs/2501.12948} {\path{arXiv:2501.12948}}.

\bibitem{dubey2024llama}
A.~Grattafiori, A.~Dubey, A.~Jauhri, A.~Pandey, A.~Kadian, A.~Al-Dahle,
  A.~Letman, A.~Mathur, A.~Schelten, A.~Vaughan, et~al., The {Llama} 3 herd of
  models (2024).
\newblock \href {http://arxiv.org/abs/2407.21783} {\path{arXiv:2407.21783}}.

\bibitem{Birhane2023Science}
A.~Birhane, A.~Kasirzadeh, D.~Leslie, S.~Wachter, Science in the age of large
  language models, Nat. Rev. Phys. 5 (2023) 277--280.

\bibitem{Azerbayev2024}
Z.~Azerbayev, H.~Schoelkopf, K.~Paster, M.~Dos~Santos, S.~McAleer, A.~Q. Jiang,
  J.~Deng, S.~Biderman, S.~Welleck, {LLEMMA}: An open language model for
  mathematics, in: International {{Conference}} on {{Learning
  Representations}}, {{ICLR}} 2024, Vienna, Austria, May 7-11, 2024.

\bibitem{Jablonka2024Leveraging}
K.~M. Jablonka, P.~Schwaller, A.~{Ortega-Guerrero}, B.~Smit, Leveraging large
  language models for predictive chemistry, Nat. Mach. Intell. 6 (2024)
  161--169.

\bibitem{lewis2020retrieval}
P.~Lewis, E.~Perez, A.~Piktus, F.~Petroni, V.~Karpukhin, N.~Goyal,
  H.~K{\"u}ttler, M.~Lewis, W.-T. Yih, T.~Rockt{\"a}schel, et~al.,
  Retrieval-augmented generation for knowledge-intensive {NLP} tasks, Adv.
  Neural Inf. Process. Syst. 33 (2020) 9459--9474.

\bibitem{chen2024metaopenfoam}
Y.~Chen, X.~Zhu, H.~Zhou, Z.~Ren, {MetaOpenFOAM}: an {LLM}-based multi-agent
  framework for {CFD} (2024).
\newblock \href {http://arxiv.org/abs/2407.21320} {\path{arXiv:2407.21320}}.

\bibitem{Chen2025MetaOpenFOAM}
Y.~Chen, X.~Zhu, H.~Zhou, Z.~Ren, {{MetaOpenFOAM}} 2.0: Large language model
  driven chain of thought for automating {{CFD}} simulation and post-processing
  (2025).
\newblock \href {http://arxiv.org/abs/2502.00498} {\path{arXiv:2502.00498}}.

\bibitem{Pandey2025OpenFOAMGPTa}
S.~Pandey, R.~Xu, W.~Wang, X.~Chu, {{OpenFOAMGPT}}: A retrieval-augmented large
  language model ({{LLM}}) agent for {{OpenFOAM-based}} computational fluid
  dynamics, Phys. Fluids 37 (2025) 35120.

\bibitem{Wang2025status}
W.~Wang, R.~Xu, J.~Feng, Q.~Zhang, X.~Chu, A status quo investigation of large
  language models towards cost-effective {{CFD}} automation with
  {{OpenFOAMGPT}}: {{ChatGPT}} vs. {{Qwen}} vs. {{Deepseek}} (2025).
\newblock \href {http://arxiv.org/abs/2504.02888} {\path{arXiv:2504.02888}}.

\bibitem{siriwardhana2023improving}
S.~Siriwardhana, R.~Weerasekera, E.~Wen, T.~Kaluarachchi, R.~Rana,
  S.~Nanayakkara, Improving the domain adaptation of retrieval augmented
  generation ({RAG}) models for open domain question answering, Trans. Assoc.
  Comput. Linguist. 11 (2023) 1--17.

\bibitem{zhang2024raft}
T.~Zhang, S.~G. Patil, N.~Jain, S.~Shen, M.~Zaharia, I.~Stoica, J.~E. Gonzalez,
  {RAFT}: Adapting language model to domain specific {RAG}, in: First
  Conference on Language Modeling, Philadelphia, PA, USA, October 7-9, 2024.

\bibitem{gao2024retrievalaugmented}
Y.~Gao, Y.~Xiong, X.~Gao, K.~Jia, J.~Pan, Y.~Bi, Y.~Dai, J.~Sun, Q.~Guo,
  M.~Wang, H.~Wang, Retrieval-augmented generation for large language models: A
  survey (2024).
\newblock \href {http://arxiv.org/abs/2312.10997} {\path{arXiv:2312.10997}}.

\bibitem{lyu2024curd}
Y.~Lyu, Z.~Li, S.~Niu, F.~Xiong, B.~Tang, W.~Wang, H.~Wu, H.~Liu, T.~Xu,
  E.~Chen, {CRUD-RAG}: A comprehensive chinese benchmark for
  retrieval-augmented generation of large language models, ACM Trans. Inf.
  Syst. 43 (2024) 1--32.

\bibitem{Ziegler2020Finetuning}
D.~M. Ziegler, N.~Stiennon, J.~Wu, T.~B. Brown, A.~Radford, D.~Amodei,
  P.~Christiano, G.~Irving, Fine-tuning language models from human preferences
  (2020).
\newblock \href {http://arxiv.org/abs/1909.08593} {\path{arXiv:1909.08593}}.

\bibitem{Ouyang2022Training}
L.~Ouyang, J.~Wu, X.~Jiang, D.~Almeida, C.~Wainwright, P.~Mishkin, C.~Zhang,
  S.~Agarwal, K.~Slama, A.~Ray, J.~Schulman, J.~Hilton, F.~Kelton, L.~Miller,
  M.~Simens, A.~Askell, P.~Welinder, P.~F. Christiano, J.~Leike, R.~Lowe,
  Training language models to follow instructions with human feedback, Adv.
  Neural Inf. Process. Syst. 35 (2022) 27730--27744.

\bibitem{Wei2021Finetuned}
J.~Wei, M.~Bosma, V.~Zhao, K.~Guu, A.~W. Yu, B.~Lester, N.~Du, A.~M. Dai, Q.~V.
  Le, Finetuned language models are zero-shot learners, in: International
  {{Conference}} on {{Learning Representations}}, {{ICLR}} 2021, Virtual, May
  3-7, 2021.

\bibitem{hu2022lora}
E.~J. Hu, Y.~Shen, P.~Wallis, Z.~Allen-Zhu, Y.~Li, S.~Wang, L.~Wang, W.~Chen,
  Lo{RA}: Low-rank adaptation of large language models, in: International
  {{Conference}} on {{Learning Representations}}, {{ICLR}} 2022, Virtual, April
  25-29, 2022.

\bibitem{wei2022chain}
J.~Wei, X.~Wang, D.~Schuurmans, M.~Bosma, F.~Xia, E.~Chi, Q.~V. Le, D.~Zhou,
  et~al., Chain-of-thought prompting elicits reasoning in large language
  models, Adv. Neural Inf. Process. Syst. 35 (2022) 24824--24837.

\bibitem{yu2024metamath}
L.~Yu, W.~Jiang, H.~Shi, J.~Yu, Z.~Liu, Y.~Zhang, J.~Kwok, Z.~Li, A.~Weller,
  W.~Liu, {MetaMath}: Bootstrap your own mathematical questions for large
  language models, in: International {{Conference}} on {{Learning
  Representations}}, {{ICLR}} 2024, Vienna, Austria, May 7-11, 2024.

\bibitem{chen2021evaluating}
M.~Chen, J.~Tworek, H.~Jun, Q.~Yuan, H.~P. D.~O. Pinto, J.~Kaplan, H.~Edwards,
  Y.~Burda, N.~Joseph, G.~Brockman, et~al., Evaluating large language models
  trained on code (2021).
\newblock \href {http://arxiv.org/abs/2107.03374} {\path{arXiv:2107.03374}}.

\bibitem{zheng-etal-2024-llamafactory}
Y.~Zheng, R.~Zhang, J.~Zhang, Y.~Ye, Z.~Luo, {L}lama{F}actory: Unified
  efficient fine-tuning of 100+ language models, in: Proceedings of the 62nd
  Annual Meeting of the Association for Computational Linguistics, Bangkok,
  Thailand, August 11-16, 2024.

\bibitem{Loshchilov2019Decoupled}
I.~Loshchilov, F.~Hutter, Decoupled weight decay regularization, in:
  International {{Conference}} on {{Learning Representations}}, {{ICLR}} 2019,
  New Orleans, LA, USA, May 6-9, 2019.

\end{thebibliography}

\end{document}